\documentclass{book}

\usepackage{amsmath,latexsym,amssymb,kcp,named,epsfig}

\newcommand{\ket}[1]{\ensuremath{|#1\rangle}}
\newcommand{\bra}[1]{\ensuremath{\langle #1|}}
\newcommand{\scprod}[2]{\ensuremath{\langle #1 | #2 \rangle }}
\newcommand{\lprod}[2]{\ensuremath{\langle #1 | #2  }}
\newcommand{\rprod}[2]{\ensuremath{ #1 | #2 \rangle }}

\newcommand{\xpect}[3]{\ensuremath{\langle #1 | #2 | #3 \rangle}}
\newcommand{\colvec}[1]{
   \left ( \begin{array}{c}
              #1_1 \\
              \vdots \\
              #1_n
              \end{array}
    \right )
}
\newcommand{\lenvec}[1]{\ensuremath{\|#1\|}}

\newcommand{\comment}[2]{

\vspace{8pt} \hrule \vspace{2pt}
\noindent {\sf \textbf{#1}: #2}
\vspace{2pt} \hrule \vspace{8pt}

}

\newcommand\remove[1]{}

\begin{document}

\paper[A Quantum Logic of Semantic Space]%
{A Quantum Logic of Down Below}%
{
  P. D. Bruza and D. Widdows and John Woods
}

\title{A Quantum Logic of Down Below}

\section{Introduction}

The logic that was purpose-built to accommodate the hoped-for
reduction of arithmetic gave to language a dominant and pivotal
place. Flowing from the founding efforts of Frege, Peirce, and
Whitehead and Russell, this was a logic that incorporated proof
theory into syntax, and in so doing made of grammar a senior
partner in the logicistic enterprise. The seniority was reinforced
by soundness and completeness metatheorems, and, in time, Quine
would quip that the ``grammar [of logic] is linguistics on
purpose"~\cite[p. 15]{Quine:1970} and that ``logic chases truth up
the tree of grammar"~\cite[p. 35]{Quine:1970}.
 Nor was the centrality of syntax lost with the G\"{o}del incompleteness results, which, except for the arithmeticization
 of syntax, would have been impossible to achieve.

Logic's preoccupation with language is no recent thing. In
Aristotle's logic of the syllogism, the target properties of
necessitation and syllogistic entailment are properties of
sentences or sets of sentences of Greek. Only with the likes of
Peirce and Frege is the rejection of natural language explicit,
each calling for a logic whose properties would attach to elements
of artificial languages, and --- after Tarski --- to such elements
in semantic relation to non-linguistic set theoretic structures.

It is hardly surprising that mathematical logic should have given such emphasis to language, given that the motivating
project of logic was to facilitate the reduction of arithmetic to an obviously analytic discipline. Still, it is also
worthy of note that the historic role of logic was to lay bare the logical structure of human reasoning. Aristotle is
clear on this point. The logic of syllogisms would serve as the theoretical core of a wholly general theory of real-life,
 two-party argumentation. Even here, the centrality of language could not be ignored. For one thing, it was obvious
 that real-life argumentation is transacted in speech. For a second, it was widely held (and still is) that reasoning is
 just soliloquial argumentation (just as argumentation is held to be reasoning in public --- out loud, so to speak). Given
 these purported equivalences, reasoning too was thought of as linguistic.

It is convenient to date the birth of modern mathematical logic
from the appearance in 1879 of Frege's great book on the language
of logic, {\em Begriffsschrift}. It is easy to think of logic as
having a relatively unfettered and richly progressive course ever
since, one in which even brutal setbacks could be celebrated as
triumphs of metalogic. There is, however, much of intervening
importance from 1904 onwards, what with developments in
intuitionist, modal, many-valued and relevant logics, which in
retrospect may seem to presage crucial developments in the second
half of that century. Suffice it here to mention Hintikka's
seminal work on epistemic logic~\cite{Hintikka:1962}, which is
notable in two important respects. One is the introduction of
agents as load-bearing objects of the logic. The other is the
influence that agents are allowed to have on what the theory
is prepared to count as its logical truths. The logical truths of
this system include its indefensible sentences, where these in
turn include sentences which it would be self-defeating for an
agent to utter (e.g., ``I can't speak a single word of English").
It is easy to see that Hintikka here allows for a sentence to be a
truth of logic if its negation is pragmatically inconsistent. To
this extent, the presence of agents in his logic occasions the
pragmaticization of its semantics.

Agents now enter logic with a certain brisk frequency. They are
either expressly there or are looming forces in theories of belief
dynamics and situation semantics, in theories of default and
non-monotonic reasoning, and in the incipient stirrings of logics
of practical reasoning. Notable as these developments are, they
all lie comfortably within the embrace of the linguistic
presumption. Agents may come or go in logic, but whether here or
there, they are, in all that makes them of interest to logicians,
manipulators of language. What is more, notwithstanding the
presence of agents, these were logics that took an interest in
human reasoning rather than human {\em reasoners}. This made for a
fateful asymmetry in which what human reasoners are like (or
should be) is read off from what human reasoning is like (or
should be).

It may be said, of course, that this is exactly the wrong way
around, that what reasoning is (or should be) can only be read off
from what reasoners are (and can be). Such a view one finds, for
example in~\cite{Gabbay:Woods:2001b}
and~\cite{Gabbay:Woods:2003b},  among logicians, and, also in the
social scientific
literature~\cite{Simon:1957,Stanovich:1999,Gigerenzer:Selten:2001a}.
Here the leading idea of the ``new logic" is twofold. First, that
logic's original mission as a theory of human reasoning should be
re-affirmed. Second, that a theory of human reasoning must take
empirical account of what human reasoners are like -- what they
are interested in and what they are capable of.

It is easy to see that the human agent is a cognitive being, that
human beings have a drive to know. They desire to know what to
believe and what to do. And since, whatever else it is, reasoning
is an aid to cognition, a theory of human reasoning must take into
account how beings like us operate as cognitive systems. Here,
too, the empirical record is indispensable. It is the first point
of contact between logic and cognition. In this way symbolic
inference becomes ``married" to computations through state (dimensional)
spaces motivated from cognition which may open the door the
large-scale operational symbolic inference systems. 
The logicians
 Barwise and Seligman have advocated such a marriage between logic and cognition
~\cite[p.234]{Book:97:Barwise:InfoFlow}. 
This
bears in an important way on what we have been calling the
linguistic presumption. For if the empirical record is anything to
go on, much of the human cognitive project is sublinguistic, and
inaccessible to introspection. This, the cognition of ``down
below", carries consequences for the new logic. If logic is to
attend to the cognizing agent, it must take the cognizer as he
comes, warts and all. Accordingly, a theory of human reasoning
must subsume a logic of down below.

These days, the logic of down below appears to have a certain memetic status. It is an idea whose time has come. In addition to the work of Gabbay and Woods, the idea is independently in play in a number of recent writings.
In~\cite{PMChurchland:1989,PMChurchland:1995} we find a connectionist approach to subconscious abductive processes (cf.~\cite{Burton:1999}).
In a series of papers, Horgan and Tiensen develop a rules without representation (RWR) framework for cognitive modeling
~\cite{Horgan:Tienson:1988,Horgan:Tienson:1989,Horgan:Tienson:1990, Horgan:Tienson:1992,Horgan:Tienson:1996,Horgan:Tienson:1999a,Horgan:Tienson:1999b} (Cf.~\cite{Guarini:2001}).
Other non-representational orientations include
~\cite{Wheeler:2001,
Sterelny:1990,Brooks:1991,Globus:1992,Shannon:1993, Thelen:Smith:1993,Wheeler:1994,Webb:1994,Beer:1995}  (Cf.~\cite{Wimsatt:1986} and~\cite{Clark:1997}).
A neural symbolic learning systemic framework is developed in~\cite{GaBrGa:2002,Garcez:Lamb:2004} and extended to abductive environments in~\cite[ch. 6]{Gabbay:Woods:2005d}.
Bruza and his colleagues advance a semantic space framework~\cite{Article:04:Bruza:Abd,Article:05:Bruza:LBD,Article:06:Bruza:Abduction}.

The present chapter is offered as a contribution to the logic of
down below. In the section to follow, we attempt to demonstrate
that the nature of human agency necessitates that there actually
be such a logic. The ensuing sections develop the suggestion that
cognition down below has a structure strikingly similar to the
physical structure of quantum states. In its general form, this is
not an idea that originates with the present authors. It is known
that there exist mathematical models from the cognitive science of
cognition down below that have certain formal similarities to
quantum mechanics. We want to take this idea seriously. We will
propose that the subspaces of von Neumann-Birkhoff lattices are
too crisp for modelling requisite cognitive aspects in relation to
subsymbolic logic. Instead, we adopt an approach which relies on projections into
nonorthogonal density states. The projection operator is motivated
from cues which probe human memory.

\section{Agency}

In this section our task is to orient the logic of down below by
giving an overview of salient features of individual cognitive
agency. Investigations of non-monotonic reasoning (NMR) have
successfully provided an impressive symbolic account of human
practical reasoning over the last two and half decades. The
symbolic characterization of practical reasoning, however, is only
part of the picture. G\"{a}rdenfors~\cite[p.
127]{Book:00:Gardenfors:ConSpace} argues that one must go under
the symbolic level of cognition. In this vein, he states, ``\dots
information about an object may be of two kinds: {\em
propositional} and {\em conceptual}. When the new information is
propositional, one learns new {\em facts} about the object, for
example, that $x$ is a penguin. When the new information is
conceptual, one {\em categorizes} the object in a new way, for
example, $x$ is {\em seen as} a penguin instead of as just a
bird''. G\"{a}rdenfors' mention of ``conceptual'' refers to the
conceptual level of a three level model of
cognition~\cite{Book:00:Gardenfors:ConSpace}.  How information is
represented varies greatly across the different levels. The
sub-conceptual level is the lowest level within which information
is carried by a connectionist representation. Within the uppermost
level information is represented symbolically. It is the
intermediate, \emph{conceptual level}, or \emph{conceptual
space}\index{space, conceptual}, which is of particular relevance
to this account. Here properties and concepts have a geometric
representation in a dimensional space.  For example, the property
of ``redness'' is represented as a convex region in a
tri-dimensional space determined by the dimensions hue,
chromaticity and brightness.  The point left dangling for the
moment is that representation at the conceptual level is rich in
associations, both explicit and implicit.  We speculate that the
dynamics of associations are primordial stimuli for practical
inferences drawn at the symbolic level of cognition.  For example,
it seems that associations and analogies generated within
conceptual space play an important role in hypothesis generation.
G\"{a}rdenfors (\cite{Book:00:Gardenfors:ConSpace}, p48) alludes
to this point when he states, ``most of scientific theorizing
takes place within the conceptual level.''

\remove{
\comment{DW}{One important link here is because some concepts are more
  ``natural'' than others. So a conceptual space version is a more
  natural conceptual scheme to give input to a Gabbay / Woods logic,
  than a Boolean set theoretic version. Essentially, set theory
  overgenerates and this gives too many possible ``concepts'' for a
  logic to be useful to a real organism or other reasoning system.

  c.f. Quine's famous `gavagai' - it could mean ``rabbits + the Eiffel
  Tower'', but any creature that stops to test this deductively
  possible interpretaton is not going to live for very long.

  There is an important relationship between the lexicon and the
  grammer / conceptual system and the reasoning system / Categories
  and Analytics going on here, and one of the pitfalls of some decades
  of research is that it has focussed on one half of this only.
}
\comment{JW}{I think that this is a point well-worth developing. If we have room
for it (we may not), I think it should go in later, perhaps as a footnote.}
}

G\"{a}rdenfors' conjecture receives strong endorsement from an
account of practical reasoning developed in
\cite{Gabbay:Woods:2003, Gabbay:Woods:2005d}, in which reasoning
on the ground is understood to function under economic
constraints. In this essay, our own point of departure is that
subsymbolic reasoning is valuable to human agents precisely for
the economies it achieves. It will help to place this assumption
in its proper context by giving a brief overview of our approach
to cognitive agency.

\subsubsection*{A Hierarchy of Agency Types}

It is useful to repeat the point that since reasoning is an aid to cognition,
a logic, when conceived of as a theory of reasoning, must take
this cognitive orientation deeply into account. Accordingly, we
will say that a {\em cognitive system} is a triple of a cognitive
agent, cognitive resources, and cognitive target performed in real
time. (See here \cite{Norman:1993,Hutchins:1995}.)
Correspondingly, a logic of a cognitive system is a principled
description of conditions under which agents deploy resources in
order to perform cognitive tasks. Such is a practical logic when
the agent it describes is a {\em practical agent}.

A practical logic is but an instance of a more general conception
of logic. The more general notion is reasoning that is
target-motivated and resource-dependent. Correspondingly, a logic
that deals with such reasoning is a Resource-Target Logic ({\em
RT}-logic). In our use of the term, a practical logic is a {\em
RT}-logic relativized to practical agents.

How agents perform is constrained in three crucial ways: in what
they are disposed towards doing or have it in mind to do (i.e.,
their {\em agendas}); in what they are capable of doing (i.e.,
their {\em competence}); and in the means they have for converting
competence into performance (i.e., their {\em resources}). Loosely
speaking, agendas here are programmes of action, exemplified by
belief-revision and belief-update, decision-making and various
kinds of case-making and criticism transacted by argument.
\footnote{Agendas are discussed at greater length in
\cite{Gabbay:Woods:2002b}.}

Agency-type is set by two complementary factors. One is the degree
of command of resources an agent needs to advance or close his (or
its) agendas. For cognitive agendas, three types of resources are
especially important.  They are (1) {\em information}, (2) {\em
time}, and (3) {\em computational capacity}. The other factor is
the height of the cognitive bar that the agent has set for
himself. Seen this way, agency-types form a hierarchy $H$
partially ordered by the relation $C$ of
commanding-greater-resources-in-support-of-higher-goals-than. {\em
H} is a poset (a partially ordered set) fixed by the ordered pair
$\langle C, X\rangle$ of the relation $C$ on the unordered set of
agents $X$.

Human agency divides roughly into the individual and the institutional. By
comparison, {\em individual} agency ranks low in $H$. For large classes of
cases, individuals
perform their cognitive tasks on the basis of less information and
less time than they might otherwise like to have, and under
limitations on the processing and manipulating of complexity. Even
so, paucity must not be confused with scarcity. There are lots of
cases in which an individual's resources are adequate for the
attainment of the attendant goal. In a rough and ready way, we can
say that the comparative modesty of an agent's cognitive goals
inoculates him against cognitive-resource scarcity. But there are
exceptions, of course.

{\em Institutional} entities contrast with human agents in all
these respects. A research group usually has more information to
work with than any individual, and more time at its disposal; and
if the team has access to the appropriate computer networks, more
fire-power than most individuals even with good PCs.  The same is
true, only more so, for agents placed higher in the hierarchy ---
for corporate actors such as NASA, and collective endeavours such
as particle physics since 1970. Similarly, the cognitive agendas
that are typical of institutional agents are by and large stricter
than the run-of-the-mill goals that motivate individual agents. In
most things, NASA aims at stable levels of scientific
confirmation, but, for individuals the defeasibly plausible often
suffices for local circumstances.

These are vital differences.  Agencies of higher rank can afford
to give maximization more of a shot. They can wait long enough to
make a try for total information, and they can run the
calculations that close their agendas both powerfully and
precisely. Individual agents stand conspicuously apart. He must do
his business with the information at hand, and, much of the time,
sooner rather than later.  Making do in a timely way with what he
knows now is not just the only chance of achieving whatever degree
of cognitive success is open to him as regards the agenda at hand;
it may also be what is needed in order to avert unwelcome
disutilities, or even death.
\remove{
\comment{JW}{Here, too I think the point well worth developing. If there is room
for it, perhaps it can be a footnote.}
\comment{We may note that higher agencies in $H$ tend to have more
  complex goals, and do not often manage the complex mapping from
  available resources to stated goals very well. In big (or small but
  badly managed) institutions, the left hand often doesn't know what
  the right hand is doing. Naturally, Boolean search engines or even
  sophisticated ranking search engines help but hit a glass ceiling
  because concepts are not well-modelled, i.e. the available
  information resources are hard to organize.}
} Given the comparative humbleness of his place in $H$, the human
individual is frequently faced with the need to practise cognitive
economies. This is certainly so when either the loftiness of his
goal or the supply of drawable resources create a cognitive
strain.  In such cases, he must turn {\em scantness} to {\em
advantage}. That is, he must (1) deal with his resource-limits and
in so doing (2) must do his best not to kill himself. There is a
tension in this dyad. The paucities with which the individual is
chronically faced are often the natural enemy of getting things
right, of producing accurate and justified answers to the
questions posed by his agenda. And yet, not only do human beings
contrive to get most of what they do right enough not to be killed
by it, they also in varying degrees prosper and flourish. 
This being so, we postulate for the individual agent {\em
slight-resource adjustment strategies} {\em (SRAS),} which he uses
to advantage in dealing with the cognitive limitations that inhere
in the paucities presently in view. We make this assumption in the
spirit of Simon \shortcite{Simon:1957} and an ensuing literature
in psychology and economics. At the heart of this approach is the
well-evidenced fact that, for ranges of cases, ``fast and frugal''
is almost as good as full optimization, and at much lower cost
\cite{Gigerenzer:Selten:2001}.  We shall not take time here to
detail the various conditions under which individuals extract
outcome economies from resource limitations and target modesty,
but the examples to follow will give some idea of how these
strategies work.

Although resource-paucity should not be equated with
resource-scarcity, it remains the case that in some sense
practical agents operate at a cognitive disadvantage. It is
advisable not to make too much of this. What should be emphasized
is that in relation to the cognitive standards that an
institutional agent might be expected to meet, the resources
available to a practical agent will typically not enable him (or
it) to achieve that standard. Whether this constitutes an
unqualified disadvantage depends on the nature of the task the
individual has set for himself and the cognitive resources
available to him. For a practical agent to suffer an unqualified
disadvantage, two factors must intersect in the appropriate way:
his resources must be inadequate for the standard he should hit,
in relation to a goal that has reasonably been set for him. So,
the measure of an agent's cognitive achievement is a function of
three factors: his cognitive {\em goal}; the {\em standard}
required (or sufficient) for achieving that goal; and the
cognitive {\em wherewithal} on which he can draw to meet that
standard.

In discharging his cognitive agendas, the practical agent tends to
set goals that he can attain and to be stocked with the
wherewithal that makes attainment possible (and frequent). In the
matter of both goals set and the execution of standards for
meeting them, the individual is a satisficer rather than an
optimizer. There are exceptions, of course; a working
mathematician won't have a solution of Fermat's Last Theorem
unless he has a full-coverage proof that is sound (and, as it
happens, extremely long).

The tendency to satisfice rather than maximize (or optimize) is
not what is {\em distinctive} of practical agency. This is a point
to emphasize. In most of what they set out to do and end up
achieving, institutional agents exhibit this same favoritism. What
matters
--- and sets them apart from the likes of us --- is not that they
routinely optimize but that they satisfice against loftier goals
and tougher standards.

\subsubsection*{Slight-resource Adjustment Strategies}

Slight-resource adjustment strategies lie at the crux of the
economy of effort, as Rescher calls it \cite[p.10]{Rescher:1996}.
They instantiate a principle of least effort, and they bear on our
tendency to minimize the expenditure of cognitive
assets.\footnote{See here the classic work of George Zipf.
\cite{Zipf:1949}}We note here some examples. 

Examples are easy to come by in discussing cognitive and economic
behaviour. More formal examples are also easy to come by, in every
field of study. Statistical studies such as opinion polls always give
results to within a given level of confidence (e.g., ``these
predictions are valid to within $\pm3\%$ with 95\% confidence''), and
part of the science of statistics lies in making reliable statements
of this nature given the size of sample taken. Medical tests are often
only correct to a known precision, and given the fequency of
false-positives, the result of a positive test-result is often a
further round of more reliable but more invasive tests.

It may be tempting to presume that such knowledge-constrained
strategies are mainly confined to empirical or practical sciences, but
this is far from the case. For example, mathematics is full of
rules-of-thumb and famous theorems that reduce difficult problems to
easy ones. These begin for many early students with the familiar division
rules, such as ``if a number ends in a 2 or a 5, it is divisible by
2 or 5'', or the more complex ``if the alternating sum of the digits
of a number is divisible by 11, the number itself is divisible by
11''. Such results do not produce the quotient of the division, but
they may tell the student whether such a computation is worth the
trouble if the goal is to end up with a whole number.
More advanced division properties are embodied in results such as {\it
  Fermat's Little Theorem}, which states that if $p$ is prime and
$1\le a \le p$, then $a^{p-1}\cong 1\ (\mathrm{mod}\ p)$. Like many important
theorems, this only gives necessary but not sufficient conditions for
a statement (in this case, the statement ``$p$ is prime'') to be
true. However, if this necessary condition holds for enough values of
$a$, we may conclude that $p$ is {\it probably} prime, which is in
fact a strong enough guarantee for some efficient encryption
algorithms. Even in mathematics, often regarded as the most exact and
uncompromising discipline, short-cuts that are close enough are not
only important, they are actively sought after.

\subsection{Hasty Generalization}

Individual cognitive agents are hasty generalizers, otherwise
known as {\em thin-slicers}. Hasty generalization is a {\em SRAS}.
In standard approaches to fallacy theory and theories of
statistical inference, hasty generalization is a blooper; it is a
serious sampling error.  This is the correct assessment if the
agent's objective is to find a sample that is guaranteed to raise
the conditional probability of the generalization, and to do so in
ways that comport with the theorems of the applied mathematics of
chance. Such is an admirable goal for agents who have the time and
know-how to construct or find samples that underwrite such
guarantees. But as J.S. Mill shrewdly observed, human individuals
often lack the wherewithal for constructing these inferences.  The
business of sample-to-generalization induction often exceeds the
resources of individuals and is better left to institutions. (See
\cite{Woods:2004}.) A related issue, even supposing that the
requisitely high inductive standards are meetable in a given
situation in which a practical agent finds himself, is whether it
is necessary or desirable for him (or it) to meet that standard.
Again, it depends on what the associated cognitive goal is. If,
for example, an individual's goal is to have a reasonable belief
about the leggedness of ocelots is, rather than to achieve the
highest available degree of scientific certainty about it, it
would suffice for him to visit the ocelot at the local zoo, and
generalize hastily "Well, I see that ocelots are four-legged".

\subsection{Generic Inference}

Often part of what is involved in a human reasoner's facility with
the one-off generalization is his tendency to eschew
generalizations in the form of universally quantified conditional
propositions. When he generalizes hastily the individual agent is
often making a {\em generic} inference.  In contrast to
universally quantified conditional propositions, a generic claim
is a claim about what is characteristically the case.  ``For all
{\em x}, if {\em x} is a ocelot, then {\em x} is four-legged'' is
one thing; ``Ocelots are four-legged'' is quite another thing
\cite{KrPeCaMeLiCh:1995}.  The first is felled by any true
negative instance, and thus is {\em brittle}. The second can
withstand multiples of true negative instances, and thus is {\em
elastic}. There are significant economies in this.  A true generic
claim can have up to lots of true negative instances. So it is
true that ocelots are four-legged, even though there are up to
lots of ocelots that aren't four-legged.  The economy of the
set-up is evident: With generic claims, it is unnecessary to pay
for every exception. One can be wrong in particular without being
wrong in general.

Generic claims are a more affordable form of generalization than
the universally quantified conditional. This is part of what
explains their dominance in the generalizations that individual
agents tend actually to make (and to get right, or some near
thing). It must not be thought, however, that what constitutes the
rightness (or some near thing) of an individual's hasty
generalizations is that when he generalizes thus he generalizes to
a generic claim. Although part of the story, the greater part of
the rightness of those hasty generalizations arises from the fact
that, in making them, an individual typically has neither set
himself, nor met, the standard of inductive strength. This,
together with our earlier remarks about validity, is telling.
Given the cognitive goals typically set by practical agents,
validity and inductive strength are typically not appropriate (or
possible) standards for their attainment. This, rather than
computational costs, is the deep reason that practical agents do
not in the main execute systems of deductive or inductive logic as
classically conceived.

\subsection{Natural Kinds}

Our adeptness with generic inference and  hasty generalization is
connected to our ability to recognize {\em natural kinds}
\cite[pp.63--95]{KrPeCaMeLiCh:1995}. Natural kinds have been the
object of much metaphysical skepticism of late \cite{Quine:1969a},
but it is a distinction that appeals to various empirical
theorists.  The basic idea is evident in concepts such as {\em
frame} \cite{Minsky:1975}, {\em prototype}
\cite{Smith:Medin:1981}, {\em script} \cite{Schank:Abelson:1977}
and {\em exemplar} \cite{Rosch:1978}. It is possible, of course,
that such are not a matter of metaphysical unity but rather of
perceptual and conceptual organization.

It goes without saying that even when the goal is comparatively
modest --- say, what might plausibly be believed about something
at hand --- not every hasty generalization that could be made
comes anywhere close to hitting even that target. The (defeasible)
rule of thumb is this: The hasty generalizations that succeed with
these more modest goals are by and large those we actually draw in
actual cognitive practice. We conjecture that the comparative
success of such generalizations is that they generalize to generic
propositions, in which the process is facilitated by the agent's
adeptness in recognizing natural kinds.
In section \ref{QL-generalization-sec}, we discuss the extent to which
a quantum logical framework provides a more useful model for
adapting to natural kinds than either Boolean set theory or taxonomy.

\subsection{Consciousness}

A further important respect in which individual agency stands
apart from institutional agency is that human agents are
conscious. (The consciousness of institutions, such as it may be
figuratively speaking, supervenes on the consciousness of the
individual agents who constitute them.) Consciousness is both a
resource and a limitation.  Consciousness has a narrow bandwidth.
This makes most of the information that is active in a human
system at a time consciously unprocessible at that time. In what
the mediaevals called the {\em sensorium} (the collective of the
five senses operating concurrently), there exist something in
excess of 10 million bits of information per second; but fewer
than 40 bits filter into consciousness at those times. Linguistic
agency involves even greater informational entropy. Conversation
has a bandwidth of about 16 bits per
second.\footnote{\cite{Zimmermann:1989}. Here is John Gray on the
same point: ``If we do not act in the way we think we do, the
reason is partly to do with the bandwidth of consciousness --- its
ability to transmit information measured in terms of bits per
second. This is much too narrow to be able to register the
information we routinely receive and act on. As organisms active
in the world, we process perhaps 14 million bits of information
per second. The bandwidth of consciousness is around eighteen
bits. This means that we have conscious access to about a
millionth of the information we daily use to survive" \cite[p.
66]{Gray:2002}.}

The narrow bandwidth of  consciousness bears on the need for
cognitive economy.  It helps elucidate what the scarcity of
information consists in.  We see it explained that at any given
time the human agent has only slight information by the fact that
if it is consciously held information there is a bandwidth
constraint which regulates its quantity. There are also devices
that regulate consciously processible information as to {\em
type}. A case in point is informational relevance.  When H.P.
Grice issued the injunction, ``Be relevant'', he left it
undiscussed whether such an imperative could in fact be honoured
or ignored by a conscious act of will. There is evidence that the
answer to this question is ``No"; that, in lot's of cases, the
mechanisms that steer us relevantly in the transaction of our
cognitive tasks, especially those that enable us to discount or
evade irrelevance, are automatic and pre-linguistic
\cite{Gabbay:Woods:2003}.  If there is marginal capacity in us to
heed Grice's maxim by consciously sorting out relevant from
irrelevant information, it is likely that these informational
relevancies are less conducive to the closing of cognitive agendas
than the relevancies that operate ``down below''. Thus vitally
relevant information often can't be processed consciously, and
much of what can is not especially vital.\footnote{Consider here
taxonomies of vision in which implicit perception has a
well-established place \cite{Rensink:2000}.}

Consciousness can claim the distinction of being one of the
toughest problems, and correspondingly, one of the most
contentious issues in the cognitive sciences.  Since the
agency-approach to logic subsumes psychological factors, it is an
issue to which the present authors fall heir, like it or not. Many
researchers accept the idea that information carries negative
entropy, that it tends to impose order on chaos.\footnote{Thus
Colin Cherry: ``In a descriptive sense, entropy is often referred
to as a `measure of disorder' and the Second Law of Thermodynamics
as stating that `systems can only proceed to a state of increased
disorder; as time passes, entropy can never decrease.'  The
properties of a gas can change only in such a way that our
knowledge of the positions and energies of the particles lessens;
randomness always increases.  In a similar descriptive way,
information is contrasted, as bringing order out of chaos.
Information, then is said to be `like' negative energy" \cite[p.
215]{Cherry:1966}.}  If true, this makes consciousness a
thermodynamically expensive state to be in, since consciousness is
a radical suppressor of information.  Against this are critics who
abjure so latitudinarian a conception of information
\cite{Hamlyn:1990} and who remind us that talk about entropy is
most assured scientifically for closed systems (and that ordinary
individual agents are hardly {\em that}).

The grudge against promiscuous ``informationalism", in which even
physics goes digital \cite{Wolfram:1984}, is that it fails to
explain the distinction between energy-to-energy transductions and
energy-to-information transformations \cite[p. 94]{Tallis:1999}.
Also targeted for criticism is the view that consciousness arises
from or inheres in neural processes.  If so, ``[h]ow does the
energy impinging on the nervous system become transformed into
consciousness?" \cite[p. 94]{Tallis:1999}.

In the interests of economy, we decline to join the metaphysical
fray over consciousness.  The remarks we have made about
consciousness are intended not as advancing the metaphysical
project but rather as helping characterize the economic
limitations under which individual cognitive agents are required
to perform.

Consciousness is tied to a family of cognitively significant
issues.  This is reflected in the less than perfect concurrence
among the following pairs of contrasts:
\begin{quote}
\begin{enumerate}
\item conscious v unconscious processing \item controlled v
automatic processing \item attentive v inattentive processing
\item voluntary v involuntary processing \item linguistic v
nonlinguistic processing \item semantic v nonsemantic processing
\item surface v depth processing
\end{enumerate}
\end{quote}
What is striking about this septet of contrasts is not that they
admit of large intersections on each side, but rather that their
concurrence is approximate at best.  For one thing, ``tasks are
never wholly automatic or attentive, and are always accomplished
by mixtures of automatic and attentive processes" \cite[p.
50]{Shiffrin:1997}.  For another, ``depth of processing does not
provide a promising vehicle for distinguishing consciousness from
unconsciousness (just as depth of processing should not be used as
a criterial attribute for distinguishing automatic processes
$\dots$" \cite[p. 58]{Shiffrin:1997}).  Indeed ``[s]ometimes
parallel processing produces an advantage for automatic
processing, but not always \ldots.  Thoughts high in consciousness
often seem serial, probably because they are associated with
language, but at other times consciousness seems parallel
$\ldots$" \cite[p. 62]{Shiffrin:1997}.

It is characteristic of agents of all types to adjust their
cognitive targets upwards as the cognitive resources for attaining
them are acquired. A practical agent may take on commitments
previously reserved for agents of higher rank if, for example, he
is given the time afforded by a tenured position in a university,
the information stored in the university's library and in his own
PC, and the fire-power of his university's mainframe. In like
fashion, institutional agents constantly seek to expand their
cognitive resources (while driving down the costs of their
acquisition, storage and deployment), so that even more demanding
targets might realistically be set. Accordingly, agents tend
toward the enhancement of cognitive assets when this makes
possible the realization of cognitive goals previously
unattainable (or unaffordable). Asset enhancement is always tied
to rising levels of cognitive ambition. In relation to cognitive
tasks adequately performed with present resources, an interest in
asset enhancement is obsessive beyond the range of what would
count as natural and proportionate improvements upon what is
already adequately dealt with.

\subsection{Subsymbolic reasoning}

Practical reasoning is reasoning performed by practical agents,
and is therefore subject to economic constraints. In this
connection, we advance the following conjecture: It may well be
that because such associations are formed below the symbolic level
of cognition, significant cognitive economy results.  This is not
only interesting from a cognitive point of view, but also opens
the door to providing a computationally tractable practical
reasoning systems, for example, operational abduction to drive
scientific discovery in biomedical
literature~\cite{Article:04:Bruza:Abd,Article:06:Bruza:Abduction}

The appeal of G\"{a}rdenfors' cognitive model is that it allows
inference to be considered not only at the symbolic level, but
also at the conceptual (geometric) level.  Inference at the
symbolic level is typically a linear, deductive process.  Within a
conceptual space, inference takes on a decidedly associational
character because associations are often based on similarity
(e.g., semantic or analogical similarity), and notions of
similarity are naturally expressed within a dimensional space. For
example, G\"{a}rdenfors states that a more natural interpretation
of ``defaults'' is to view them as ``relations between
concepts''.\footnote{In the theory of Gabbay and Woods, default
reasoning is a core slight-resource compensation strategy.} This
is a view which flows into the account which follows: the strength
of associations between concepts change dynamically under the
influence of context. This, in turn, influences the defaults
haboured within the symbolic level of cognition.

It is important to note the paucity of representation at the
symbolic level and reflect how symbolic reasoning systems are
hamstrung as a result. In this connection, G\"ardenfors (\cite[p.
127]{Book:00:Gardenfors:ConSpace}) states, `` ..information about
categorization can be quite naturally transfered to propositional
information: categorizing $x$ as an emu, for example, can be
expressed by the proposition ``$x$ is an emu''. This
transformation into the propositional form, however, tends to
suppress the internal {\em structure} of concepts. Once one
formalizes categorizations of objects by {\em predicates} in a
first-order language, there is a strong tendency to view the
predicates as primitive, atomic notions and to forget that there
are rich relations among concepts that disappear when put into
standard logical formalism.''

The above contrast between the conceptual and symbolic levels raises
the question as to what are the implications for providing an account
of practical reasoning.  G\"{a}rdenfors states that concepts generate
``expectations that result in different forms of \emph{non-monotonic
reasoning}'', which are summarized as follows:

\subsection*{Change from a general category to a subordinate}

When shifting from a basic category, e.g., ``bird" to a subordinate
category, e.g., ``penguin", certain default associations are given up
(e.g., ``Tweety flies''), and new default properties may arise (e.g.,
``Tweety lives in Antarctica'').

\subsection*{Context effects}

The context of a concept triggers different associations that ``lead
to non-monotonic inferences''. For example, {\em Reagan} has default
associations ``Reagan is a president'', ``Reagan is a republican'' etc.,
but {\em Reagan} seen in the context of {\em Iran} triggers associations of
``Reagan'' with ``arms scandal'', etc.

\subsection*{The effect of contrast classes}

Properties can be relative, for example, ``a tall Chihuahua is not
a tall dog'' (\cite[p. 119]{Book:00:Gardenfors:ConSpace}),. In the
first contrast class ``tall'' is applied to Chihuahuas and the
second instance it is applied to dogs in general.  Contrast
classes generate conceptual subspaces, for example, skin colours
form a subspace of the space generated by colours in general.
Embedding into a subspace produces non-monotonic effects.  For
example, from the fact that $x$ is a white wine and also an
object, one cannot conclude that $x$ is a white object (as it is
yellow).

\subsection*{Concept combination}

Combining concepts results in non-monotonic effects.  For example,
\emph{metaphors} (\cite[p. 130]{Book:00:Gardenfors:ConSpace}),
Knowing that something is a lion usually leads to inferences of
the form that it is alive, that it has fur, and so forth. In the
combination, {\em stone lion}, however, the only aspect of the
object that is lion-like is its shape. One cannot conclude that a
stone lion has the other usual properties of a lion, and thus we
see the non-monotonicity of the combined concept.

An example of the non-monotonic effects of concept combination not involving
metaphor is the following:
\emph{
A guppy
is not a typical pet, nor is guppy is a typical fish, but a guppy is a
typical pet fish}.

In short, concept combination leads to conceptual change. These
correspond to revisions of the concept and parallel belief revisions
modelled at the symbolic level, the latter having received thorough
examination in the artificial intelligence literature.

The preceding characterization of the dynamics of concepts and
associated non-monotonic effects is intended to leave the
impression that a lot of what happens in connection with practical
reasoning takes place within a conceptual (geometric) space, or a
space of down-below. What is more, this impression may provide a
foothold towards realizing genuine operational systems. This would
require that at least three issues be addressed. The first is that
a computational variant of the conceptual level of cognition is
necessary. Secondly, the non-monotonic effects surrounding
concepts would need to be formalized and implemented. Thirdly, the
connection between these effects and NMR at the symbolic level
needs to be specified. This account will cover aspects related to
the first two of these questions. Computational approximations of
conceptual space will be furnished by semantic space models which
are emerging from the fields of cognition and computational
linguistics.  Semantic space models not only provide a cognitively
motivated basis to underpin human practical reasoning, but from a
mathematical perspective, they show a marked similarity with
quantum mechanics (QM) \cite{Article:04:Aerts:QM}. This introduces
the tantalizing and unavoidably speculative prospect of
formalizing aspects of human practical reasoning via QM.

\section[Semantic Space]{
  Semantic space: computational approximations of conceptual space
}

To illustrate how the gap between cognitive knowledge representation
and actual computational representations may be bridged, the Hyperspace Analogue to
Language (HAL)\index{hyperspace analogue to language} semantic space model is
employed~\cite{Article:96:Lund:HAL,Article:98:Burgess:HAL}.  HAL
produces representations of words in a high dimensional space that
seem to correlate with the equivalent human representations.  For
example, ``...simulations using HAL accounted for a variety of
semantic and associative word priming effects that can be found in the
literature...and shed light on the nature of the word relations found
in human word-association norm data''\cite{Article:96:Lund:HAL}.
Given an $n$-word vocabulary, HAL computes an $n \times n$ matrix
constructed by moving a window of length $l$ over the corpus by one
word increment ignoring punctuation, sentence and paragraph
boundaries.  All words within the window are considered as
co-occurring with the last word in the window with a strength
inversely proportional to the distance between the words. Each row
$i$ in the matrix represents accumulated weighted associations of word
$i$ with respect to other words which preceded $i$ in a context
window.  Conversely, column $i$ represents accumulated weighted
associations with words that appeared after $i$ in a window.  For
example, consider the text ``President Reagan ignorant of the arms
scandal'', with $l=5$, the resulting HAL matrix $H$ would be:
\begin{table}[h!]
\begin{center}
\begin{tabular}{|c|c|c|c|c|c|c|c|} \hline
 & arms & ig & of & pres &  reag & scand & the \\ \hline
 arms & 0 & 3 & 4 & 1& 2& 0 & 5 \\ \hline
 ig & 0 & 0 & 0 & 4 & 5 & 0 & 0 \\ \hline
 of & 0 & 5 & 0 & 3 & 4 & 0 & 0 \\ \hline
 pres & 0 & 0 & 0 & 0 & 0 & 0 & 0 \\ \hline
 reag & 0 & 0 & 0 & 5 & 0 & 0 & 0 \\ \hline
 scand & 5 & 2 & 3 & 0 & 1 & 0 & 4 \\ \hline
the & 0 & 4& 5 & 2 & 3 & 0 & 0 \\ \hline
\end{tabular}
\caption{A simple semantic space computed by HAL}
\end{center}
\label{table:reaganeg}
\end{table}
\begin{figure}
\begin{center} \begin{minipage}{10cm}{\small\tt
\begin{tabbing}
def calculate\_hal(documents, n) \\
\hspace{1em}\=  HAL = 2DArray.new() \\
\>              for d in documents \{ \\
\>\hspace{1em}\=    for i in 1 .. d.len \{ \\
\>\>\hspace{1em}\=      for j in max(1,i-n) .. i-1 \{ \\
\>\>\>\hspace{1em}\=        HAL[d.word(i),d.word(j)] += n+1-(i-j) \\
\>  \}\}\} \\
\>  return HAL \\
end
\end{tabbing}}\end{minipage}
\end{center}
\caption{
  Algorithm to compute the HAL matrix for a collection of
  documents. It is assumed that the documents have been pruned of stop
  words and punctuation.
}
\end{figure}

If word precedence information is considered unimportant the matrix
$S=H+H^T$ denotes a symmetric matrix in which $S[i,j]$ reflects the
strength of association of word $i$ seen in the context of word $j$,
irrespective of whether word $i$ appeared before or after word $j$ in
the context window.  The column vector $S_j$ represents the strengths
of association between $j$ and other words seen in the context of the
sliding window: the higher the weight of a word, the more it has
lexically co-occurred with $j$ in the same context(s).  For example,
table~\ref{table:reagan} illustrates the vector representation for
``Reagan'' taken from a matrix $S$ computed from a corpus of 21578
Reuters\footnote{The Reuters-21578 collection is standard test collection used for research into automatic text classification} news feeds taken from the year 1988.
(The weights in the table are not normalized). Highly weighted associations reflect Reagan in his presidential role dealing with congress, tax, vetoes etc. In addition, the more highly weighted association reflect a default-like character, e.g., ``president" and ``administration". Associations with lower weights seem to reflect the trade war with Japan (``japan", ``tariffs") and the Iran-contra scandal (``Iran", ``arms").
In other words, the representation of Reagan represents a mixture of different ``senses" of Reagan. This facet is intuitively similar to the QM phenomenon of a particle being in a state of superposition.
\begin{table}[t]
\begin{center}
\begin{tabular}{|p{11cm}|} \hline
president (5259), administration (2859), trade (1451), house (1426),
budget (1023), congress (991), bill (889), tax (795), veto (786),
white (779), japan (767), senate (726), iran (687), billion (666),
dlrs (615), japanese (597), officials (554), arms (547), tariffs
(536) \dots \\ \hline
\end{tabular}
\end{center}
\caption{Example representation of the word ``Reagan''}
\label{table:reagan}
\end{table}

HAL is an exemplar of a growing ensemble of computational models
emerging from cognitive science, which are generally referred to
as {\em semantic
spaces}~\cite{Article:96:Lund:HAL,Article:98:Burgess:HAL,%
Article:00:Lowe:SemanticSpace,Article:01:Lowe:SemanticSpace,%
Article:97:Landauer:LSA,Article:98:Landauer:LSA,%
Article:97:Patel:SemanticSpace,Article:98:Schuetze:SemanticSpace,%
Article:99:Levy:SemanticSpace,Article:02:Sahlgren:SemanticSpace}.
Even though there is ongoing debate about specific details of the
respective models, they all feature a remarkable level of
compatibility with a variety of human information processing tasks
such as word association.  Semantic spaces\index{space, semantic}
provide a geometric, rather than propositional, representation of
knowledge.  They can be considered to be approximations of
conceptual space proposed by
G\"{a}rdenfors~\cite{Book:00:Gardenfors:ConSpace}, and of
reasoning down below as proposed
by~\cite{Gabbay:Woods:2003,Gabbay:Woods:2005d}.

Within a conceptual space, knowledge has a dimensional structure.  For
example, the property colour can be represented in terms of three
dimensions: hue, chromaticity, and brightness.
G\"{a}rdenfors argues that a
property is represented as a convex region in a geometric space.  In
terms of the example, the property ``red'' is a convex region within
the tri-dimensional space made up of hue, chromaticity and brightness.
The property ``blue'' would occupy a different region of this space.
A domain is a set of integral dimensions in the sense that a value in
one dimension(s) determines or affects the value in another
dimension(s).  For example, the three dimensions defining the colour
space are integral since the brightness of a colour will affect both
its saturation (chromaticity) and hue.  G\"{a}rdenfors extends the
notion of properties into concepts, which are based on domains.  The
concept ``apple'' may have domains taste, shape, colour, etc.  Context
is modelled as a weighting function on the domains, for example, when
eating an apple, the taste domain will be prominent, but when playing
with it, the shape domain will be heavily weighted (i.e., it's
roundness).  One of the goals of this article is to provide both a
formal and operational account of this weighting function.

Observe the distinction between representations at the symbolic
and conceptual levels. At the symbolic level ``apple'' can be
represented as the atomic proposition $apple(x)$. However, within
a conceptual space (conceptual level), it has a representation
involving multiple inter-related dimensions and domains.
Colloquially speaking, the token ``apple'' (symbolic level) is the
tip of an iceberg with a rich underlying representation at the
conceptual level.  G\"{a}rdenfors points out that the symbolic and
conceptual representations of information are not in conflict with
each other, but are to be seen as ``different perspectives on how
information is described''.

Barwise and Seligman~\cite{Book:97:Barwise:InfoFlow} also propose a
geometric foundation to their account of inferential information
content via the use of real-valued state spaces\index{space, state
space}.  In a state space, the colour ``red'' would be represented as
a point in a tri-dimensional real-valued space.  For example,
brightness can be modelled as a real-value between white (0) and black
(1).  Integral dimensions are modelled by so called observation
functions defining how the value(s) in dimension(s) determine the
value in another dimension.  Observe that this is a similar proposal,
albeit more primitive, to that of G\"{a}rdenfors as the
representations correspond to points rather than regions in the space.

Semantic space models are also an approximation of Barwise and
Seligman state spaces whereby the dimensions of the space
correspond to words. A word $j$ is a point in the space.  This
point represents the ``state'' in the context of the associated
text collection from which the semantic space was computed.  If
the collection changes, the state of the word may also change.
Semantic space models, however, do not make provision for integral
dimensions.  An important intuition for the following is the state
of a word in semantic space is tied very much with its ``meaning",
and this meaning is context-sensitive. Further,
context-sensitivity will be realized by state changes of a word.

In short, HAL, and more generally semantic spaces, are a
promising, pragmatic means for knowledge representation based on
text.  They are computational approximations, albeit rather
primitively, of G\"{a}rdenfors' conceptual space.  Moreover, due
to their cognitive track record, semantic spaces would seem to be
a fitting foundation for considering realizing computational
variants of human reasoning. Finally, it has been shown that a semantic space is formally a density matrix, a notion from QM~\cite{Article:04:Aerts:QM,Article:05:Bruza:QM}.
This opens the door to exploring further connections with QM.

\section{Bridging Semantic Space and Quantum Mechanics}

HAL exemplifies how a semantic space model assigns each word in a given vocabulary a point in a finite dimensional vector space.
Lowe~\cite{Article:01:Lowe:SemanticSpace} formalizes semantic space models as a quadruple $\langle A, B, F, M \rangle$ where
\begin{itemize}
\item $B$ is a set of $m$ basis elements 
\item $A$ is a function which maps the co-occurrence frequencies between words in a vocabulary $V$ and the basis elements so each $w \in V$ is represented by a vector $(A(b_1,w), \ldots, A(b_m,w))$
\item $F$ is a function which maps pairs of vectors onto continuous valued quantity. The interpretation of $F$ is often ``semantic similarity" between the two vectors in question.
\item $M$ is a transformation which takes one semantic space and maps it into another, for example via dimensional reduction
\end{itemize}
A semantic space\footnote{Bear in mind that the term ``space" should not be interpreted as a ``vector space". This unfortunate blurring between ``matrix" and ``space" in the technical sense occurs because ``semantic space" is a term from the cognitive science literature.}
 $S$ is an instance of the range of the function $A$.
That is, $S$ is a $m \times n$ matrix where the columns $\{1,
\ldots, n\}$ correspond to vector representations of words.A typical
method for deriving the vocabulary $V$ is to tokenize the 
corpus from which the semantic space is computed and remove non information bearing words such as ``the'',
``a'', etc.  The letters $u, v, w$ will be used to identify individual
words.

The interpretation of the basis elements corresponding to the rows $\{1 \ldots m\}$ depends of the type of
semantic space in question.  For example, table~\ref{table:reaganeg}
illustrates that HAL produces a square matrix in which the rows are
also interpreted as representations of terms from the vocabulary $V$.  
In contrast, a row in
the semantic space models produced by\index{latent semantic analysis}
Latent Semantic Analysis~\cite{Article:98:Landauer:LSA} corresponds to
a text item, for example, a whole document, a paragraph, or even a
fixed window of text, as above. The value $S[t, w]=x$ denotes the
salience $x$ of word $w$ in text $t$. Information-theoretic approaches
are sometimes use to compute salience. Alternatively, the (normalized) frequency of
word $w$ in context $t$ can be used.

For reasons of a more straightforward embedding of semantic space into
QM, we will focus on square, symmetric semantic spaces ($m=n$).
The following draws from~\cite{Book:04:Rijsbergen:QM}

A word $w$ is represented as a column vector in $S$:
\begin{equation} \quad \quad \quad \quad
\ket{w} = \colvec{w}
\end{equation}
The notation on the LHS is called a \emph{ket}, and originates
from quantum physicist Paul Dirac.  Conversely, a row vector $v =
(v_1, \ldots, v_n)$ is denoted by the \emph{bra} \bra{v}.

\index{ket}
\index{bra}

Multiplying a ket by a scalar $\alpha$ is as would be expected:
\begin{equation} \quad \quad \quad
 \alpha\ket{w} =
 \left ( \begin{array}{c}
             \alpha w_1 \\
              \vdots \\
              \alpha w_n
             \end{array}
    \right )
\end{equation}
Addition of vectors $\ket{u} + \ket{v}$ is also as one would expect.
In Dirac notation, the scalar product of two $n$-dimensional
real\footnote{QM is founded on complex vector spaces. We restrict our
attention to finite vector spaces of real numbers.} valued vectors $u$
and $v$ produces a real number:
\begin{equation} \quad \quad \quad
\scprod{u}{v} = \sum_{i=1}^n u_i v_i
\end{equation}

The outer product \ket{u}\bra{u} produces a $n \times  n$ symmetric matrix.  Vectors $u$
and $v$ are \emph{orthogonal} iff $\scprod{u}{v} = 0$.  Scalar product
allows the length of a vector to be defined: $\lenvec{u} =
\sqrt{\scprod{u}{u}}$.  A vector \ket{u} can be normalized to unit
length ($\lenvec{u}=1$) by dividing each of its components by the
vector's length: $\frac{1}{\lenvec{u}}\ket{u}$.

\index{space, Hilbert}
A Hilbert space is a complete\footnote{The notion of a ``complete''
vector space should not be confused with ``completeness'' in
logic. The definition of a completeness in a vector space is rather
technical, the details of which are not relevant to this account.}
inner product space.  In the formalization to be presented in ensuing
sections, a semantic space $S$ is an $n$-dimensional real-valued
Hilbert space using Euclidean scalar product as the inner product.

A Hilbert space allows the state of a quantum system to
be represented.
It is important to note that a Hilbert
space is an \emph{abstract} state space meaning QM does not prescribe
 the state space of specific systems such as electrons. This is the
responsibility of a physical theory such as quantum electrodynamics.
Accordingly, it is the responsibility of semantic space theory to
offer the specifics: In a nutshell, a ket \ket{w} describes the state of ``meaning"
of a word $w$. It is akin to a particle in QM.
The state of a word changes due to context effects in
a process somewhat akin to quantum collapse\index{quantum collapse}.
This in turn bears on practical inferences drawn due to context effects of word seen
together with other words as described above.

In QM, the state can represent a superposition of potentialities.  By
way of illustration consider the state $\sigma$ of a quantum bit, or
\emph{qubit} as:
\begin{equation} \quad \quad \quad
\ket{\sigma} = \alpha\ket{0} + \beta\ket{1}
\end{equation}
where $\alpha^2 + \beta^2 = 1$.  The vectors \ket{0} and \ket{1}
represent the characteristic states, or \emph{eigenstates}, of ``off'' and
``on''. Eigenstates \index{eigenstate} are sometimes referred to as \emph{pure},
or \emph{basis} states. They can be pictured as defining orthogonal axes in a 2-
D plane:
\begin{equation} \quad \quad \quad
 \alpha\ket{0} =
 \left ( \begin{array}{c}
              0 \\
               1
             \end{array}
    \right )
\end{equation}
and
\begin{equation} \quad \quad \quad
 \alpha\ket{1} =
 \left ( \begin{array}{c}
              1 \\
               0
             \end{array}
    \right )
\end{equation}
The state $\sigma$ is a linear combination of eigenstates. Hard though
it is to conceptualize, the linear combination allows the state of the
qubit to be a mixture of the eigenstates of being ``off'' and
``on'' at the same time.

In summary, a quantum state\index{quantum state} encodes the
probabilities of its measurable properties, or eigenstates. The
probability of observing the qubit being off (i.e., \ket{0} is
$\alpha^2$). Similarly, $\beta^2$ is the probability of observing it
being ``on''.

The above detour into QM raises questions in relation to semantic
space. What does it mean that a word is a superposition - a ``mixture
of potentialities''? What are the eigenstates of a word?

\subsection{Mixed and eigenstates of a word meaning}.

\remove{
In order to provide some intuition about how QM relates to semantic space,
consider the word ``suit".
In isolation it is ambiguous - it may refer refer to an item of clothing, or a
legal procedure. However, when seen in the context of the word ``grey", the
ambiguity resolves into the sense of the word dealing with clothing. The
connection with QM is the following. The ``meanings" of words in semantic space
are represented as vectors which comprise mixtures of potential senses. This is
akin to a quantum particle in superposition.
When a measurement is performed, the particle collapses onto an eigenstate.
Recent work has shown how the collapse of word meanings onto a sense parallels
the collapse of a quantum
particle~\cite{Article:05:Bruza:QM,Article:05:Aerts:QMa,Article:03:Widdows:QM}.
It is important to stress how the issue of context has been modelled as a
``measurement"- seeing a word in the contexts of other words acts like a
measurement. }

Consider the following traces of text from the Reuters-21578
collection: \emph{
\begin{itemize}
\item President Reagan was ignorant about much of the Iran arms scandal
\item Reagan says U.S to offer missile treaty
\item Reagan seeks more aid for Central America
\item Kemp urges Reagan to oppose stock tax.
\end{itemize}
}

Each of these is a window which HAL will process accumulating weighted
word associations in relation to the word ``Reagan''. In other
words, included in the HAL vector for ``Reagan'' are associations
dealing with the Iran-contra scandal, missile treaty negotiations with
the Soviets, stock tax etc.  The point is when HAL runs over the full
collection, the vector representation for
``Reagan''is a mixture of eigenstates, whereby an eigenstate corresponds to a
particular ``sense", or ``characteristic meaning" of the concept ``Reagan". For
example, Reagan, in the political sense, in the sense dealing with the Iran-Contra scandal, etc.
The senses of a concept are equivalent of the eigenstates of a particle in QM~\cite{Article:05:Bruza:QM,Article:05:Aerts:QMa,Article:03:Widdows:QM}.

Consider once again the
HAL matrix $H$ computed from the text ``President Reagan ignorant of
the arms scandal''. As mentioned before, $S=H+H^T$ is a real symmetric
matrix.
Consider a set of $y$ text windows
of length $l$ which are centred around a word $w$.
Associated with each such text window $j, 1 \leq j \leq m$,
is a semantic space $S_j$.  It is assumed that the semantic space is
$n$-dimensional, whereby the $n$ dimensions correspond to a fixed
vocabulary $V$ as above.  The semantic space around word $w$, denoted
by $S_w$, can be calculated by the sum:
\begin{equation} \quad \quad \quad
  S_w =\sum_{j=1}^yS_j
  \label{eqn:semspace}
\end{equation}
The above formula provides a toehold for computing a semantic space in terms of a sum of semantic spaces; each constituent semantic space corresponding to a specific sense of the concept $w$.
By way of illustration, Let the concept $w$ be ``Reagan" and assume there are a
total of $y$ traces centred on the word ``Reagan", $x$ of which  deal with the
Iran-contra issue.
These $x$ traces can be used to construct a semantic
space using equation~\ref{eqn:semspace}.
Let $S_i$ denote this semantic space.
Its associated probability  $p_i =
\frac{x}{y}$.
Assume the concept $w$ has $m$ senses. As each sense $i$
represents a particular state of $w$, each can be represented as a semantic space $S_i$ with an associated probability.

  \begin{equation} \quad \quad \quad
    S_w  = p_1S_1 + \ldots + p_mS_m
   \label{eqn:psem}
  \end{equation}
  where $p_1 + \ldots + p_m = 1$.

This formula expresses that the semantic space around a concept
$w$ can be conceived of as a linear combination of semantic
spaces around senses of $w$. The formula is intuitively close to
an analogous formula from QM whereby a density matrix can be
expressed as a probability mixture of density matrices \cite[p.
778]{Article:03:Nielsen:QM}. A density matrix represents the state
of a quantum system. A density matrix $\rho$ represents a basis
state if it has a single eigenvector (eigenstate)$\ket{e}$ whereby
$\rho = \ket{e}\bra{e}$. The vector $e$ is termed a \emph{state
vector} and is assumed to be normalized to unit length, i.e.,
$\scprod{e}{e} =1$. Otherwise a density matrix $\rho$ represents a
``mixed" state (superposition). A density matrix corresponding to
a mixed state can be expressed as a weighted combination of
density matrices corresponding to basis states. There is no
requirement that the state vectors of the pure states are
orthogonal to one another. This is a very important point.
Intuitively, it is unrealistic to require  the senses of a
concept to be orthogonal. For this reason, the term ``sense" will be
used to denote the basis state of a word meaning, rather than
``eigenstate", because, in QM, eigenstates are assumed to be
mutually orthogonal.

The connection between the notions of semantic space and density matrix have been detailed elsewhere~\cite{Article:04:Aerts:QM, Article:05:Bruza:QM}.
As mentioned in the introduction, there are various semantic space models presented in the literature. Each will involve a different rendering as density matrix.
The method adopted in this account rests on the intuition  the ket $\ket{e_i}$ in each semantic space $S_i$ of equation~\ref{eqn:psem} corresponds to a state vector representing a sense of concept $w$.
A density matrix $\rho_i$ can be formed by the product $\ket{e_i} \bra{e_i}$.
Building on this, a density matrix $\rho_w$ corresponding to the semantic space $S_w$ can be constructed as follows.
  \begin{equation} \quad \quad \quad
    \rho_w  = p_1\rho_1 + \ldots + p_m\rho_m
   \label{eqn:density}
  \end{equation}
Importantly, no assumption of orthogonality has been made.

This approach to representing a semantic space in a state contrasts approaches using the spectral decomposition of the semantic space~\cite{Article:04:Aerts:QM,Article:05:Aerts:QMII}.
As the semantic space $S_w$ is a symmetric matrix, the spectral decomposition\index{decomposition, spectral} of SVD
allows $S_w$ to be reconstructed, where $k \leq n$:
\begin{eqnarray*}
  S_w &= &\sum_{i=1}^k\ket{e_i}d_i\bra{e_i} \\
           & = & \sum_{i=1}^k d_i \ket{e_i}\bra{e_i} \\
           & = & d_1\ket{e_1}\bra{e_1} + \ldots + d_k \ket{e_k}\bra{e_k}
\end{eqnarray*}
This equation parallels the one given in equation~\ref{eqn:density}.
The eigenvalues $d_i$ relate to the probabilities of the
associated eigenvectors (eigenstates in QM terminology). Each eigenstate $\ket{e_i}$ contributes to the linear combination via the density matrix
$\ket{e_i}\bra{e_i}$. 
The eigenstates
\ket{e_i} of $S_w$ should ideally correspond to the senses of word $w$.
Unfortunately, this does not bear out in practice.
A fundamental problem is that the eigenstates $\ket{e_i}$ computed by SVD are orthogonal, and in reality the senses of a word $w$ need not be.
(See~\cite{Article:05:Bruza:QM} for more details).

\subsection{The collapse of meaning in the light of context}

We continue by connecting the above development of
quantum mechanics in semantic space to G\"{a}rdenfors' views on
the interaction of context and the meaning of concepts. He states,
``The starting point is that, for some concepts, the meaning of
the concept is determined by the \emph{context} in which it
occurs" \cite[p.119]{Book:00:Gardenfors:ConSpace}. Context effects
manifest in relation to contrast classes. In the introduction, the
Chihuahua showed how property tall is relative,  ``a tall
Chihuahua is not a tall dog''. He also illustrates how contrast
classes manifest in word combinations. Consider, ``red" in the
following combinations, ``red book", ``red wine", ``red hair",
``red skin", ``red soil". G\"{a}rdenfors argues contrast classes
generate conceptual subspaces, for example, skin colours form a
subspace of the space generated by colours in general. In other
words, each of the combinations involving ``red" results in a
separate subspace representing the particular quality of ``red",
for example, the quality of ``red" would actually be ``purple"
when ``red" is seen in the context of ``wine".

\remove{
The phenomenon of quantum collapse of meaning is closely tied to
G\"{a}rdenfors' intuitions in the following way. A concept $w$ is represented as
a semantic space $S_w$. Context acts like a quantum measurement which projects $S_w$ into a (sub)space.
In some cases the resulting space will be a subspace, like in the Iran-contra subspace of the Reagan space, but it need not be.
Consider the effect of ``stone" on the concept ``lion". In this case, the resulting space will be largely outside of the  semantic space associated with ``lion".
}

The collapse of word meaning can be thought of in terms of the quantum collapse of the particle but with an important difference: The collapse due to context may not always result in a basis state because the context may not be sufficient to fully resolve the sense in question.
By way of illustration, consider ``Reagan" in the context of ``Iran".
For the purposes of discussion, assume there are two possible senses. The first deals with the Iran-contra scandal, and the other deals with hostage crisis at the American embassy in Teheran.
The distinction between a measurement due to context and a physical measurement possibly has its roots in human memory.
Matrix models of human memory also contain the notion of superimposed memory states, and it has been argued, ``The superposition of memory traces in a vector bundle resulting from a  memory retrieval has often been considered to be a noisy signal that needs to be `cleaned up' [i.e., full collapse onto a basis state as in QM]. The point we make here is that this is \emph{not necessarily so} and that the superposition of vectors [after retrieval] is a powerful process that adds to the flexibility of memory processes." (Emphasis ours)~\cite{Wiles:Halford:1994}.

This distinction requires a less stringent notion of collapse as maintained within QM.
Consider a concept $w$ considered in the light of some context, for example, other words. The context is denoted generically by $X$.
The effect of context $X$ is brought about by a projection operator $P_x$.
Assuming the density matrix $\rho_w$ corresponding to a concept $w$, the collapse of meaning in the light of context $X$ is characterized by the following equation:
\begin{equation} \quad \quad \quad
    P_x\rho_w  = p\rho_w^x
   \label{eqn:collapse}
 \end{equation}
  where $p$ denotes the probability of collapse and $\rho_w^x$ is the state of $w$ after the ``collapse" of its meaning. 

In terms of QM, $\rho_w$ is an ``observable" meaning
an observable physical quantity. An observable is represented by a self-adjoint operator.
As $\rho_w$ is a real symmetric matrix, it is therefore also a self-adjoint operator on Hilbert space. (Recall that semantic space is a Hilbert space). This is consistent with the second axiom of QM~\cite{Byron:Fuller:1992}.
Even though this equation has the form of an eigenvalue problem, the value $p$ is not an eigenvalue.  It is a theorem that the eigenstates of a self adjoint operator belonging to different eigenvalues must be orthogonal, a requirement which is too strong for word meanings as was motivated earlier. Nevertheless,  it will be be shown later that $p$ derives from the geometry of the space as do eigenvalues.

The previous equation is also consistent with the third axiom of QM as the result of ``measurements of the observable [$\rho_w$]" is an element of ``the spectrum of the operator".
In our case, the spectrum is specified by the probability mixture given in equation~\ref{eqn:density}, but more of the flexibility of this equation is exploited than is the case in QM.
The key to this flexibility revolves around the fact that the sum of density matrices is a density matrix.
By way of illustration, equation~\ref{eqn:density} can be equivalently written as the probability mixture:
\begin{eqnarray*}
\rho_w & = p_1\rho_1 + p_2\rho_2
\end{eqnarray*}
where $p_1 + p_2 = 1$.
Let $\rho_1$ correspond to the state of ``Reagan" in the context of ``Iran" and $\rho_2$ the state of ``Reagan" in all other contexts.
Assume, that ``Reagan" is seen in the context of ``Iran".
The projection operator $P_x$ collapses $\rho_w$ onto $\rho_1$ with probability $p_1$.
Unlike, QM, the state $\rho_1$ is not a basis state but corresponds to a partially resolved sense.
Let the Iran-contra sense be denoted $\ket{c}$ and the Iranian embassy hostage crisis be denoted $\ket{h}$.
In the light of this example, the density matrix corresponding to the state after collapse due to ``Iran" would be of the form $\rho_1 = p_c\ket{c}\bra{c} + p_h\ket{h}\bra{h}$, where $p_c + p_h =1$.

It has been argued in~\cite{Article:05:Bruza:QM} that in terms of
this running example many would assume the ``Iran-contra" sense of
``Reagan" when ``Reagan" is seen in the context of ``Iran". This
phenomenon may have its roots in cognitive economy. Full
resolution requires processing, and to avoid this processing,
humans ``guess" the more likely sense (In the example, $p_c$
happens to be substantially greater than $p_h$). In other words,
we cautiously put forward the conjecture that collapse of meaning
and abductive processes go hand in hand to fully resolve the sense, i.e., collapse onto a basis state.
Even though ``full" collapse eventually results, the process is
not direct as is the the case of the collapse in QM.\footnote{For
a more detailed discussion of how the logic of abduction engages
with the cognitive economy of practical agency, see
\cite{Gabbay:Woods:2005d}. For the link between abduction and
semantic space, see \cite{Article:06:Bruza:Abduction}}.

The running example reveals
something of the nature of the projection operator $P_x$. If $P_x$
is orthogonal to a sense $\ket{e_i}$ represented by the density
matrix $\rho_i= \ket{e_i}\bra{e_i}$, then $P_x$ projects this
sense onto the zero vector $\ket{0}$. (Note the corresponding
density matrix is $\ket{0}\bra{0}$). If the projection $P_x$ is
not orthogonal to a sense $\ket{e_i}$, then it has the effect of
retrieving those senses out of the combination expressed in
equation~\ref{eqn:density}. This is not unlike the notion of a cue
which probes human memory. Cues can be used to access memory in
two ways; via \emph{matching} or \emph{retrieval}
processes. Matching entails the ``comparison
of the test cue(s) with the information stored in memory"~\cite[p
41.]{Humphreys:Bain:Burt:89}. This process measures the similarity
of the cue(s) and the memory representation. The output of this
process is a scalar quantity (i.e., a single numeric value
representing the degree or strength of the match). Memory tasks
which utilise this access procedure include recognition and
familiarity tasks. Retrieval involves the ``recovery of
qualitative information associated with a cue"~\cite[p
141.]{Humphreys:Bain:Burt:89}. This information is modelled as a
vector of feature weights. Retrieval tasks include free recall,
cued-recall, and indirect production tasks.

The intuition we will attempt to develop is that collapse of word meaning due to context is akin to a cued-recall retrieval operation driven by the projector $P_x$ on a given density matrix corresponding to the state of a word meaning. The probability of collapse $p$ is a function of the scalar quantity resulting from matching. 

In the matrix model of memory~\cite{Humphreys:Bain:Burt:89} , memory representations can include items, contexts or, combinations of items and contexts (associations).
Items can comprise stimuli, words, or concepts. Each item is modelled as a vector of feature weights. Feature weights are used to specify the degree to which certain features form part of an item. There are two possible levels of vector representation for items. These include:
\begin{itemize}
\item modality specific peripheral representations (e.g., graphemic or phonemic representations of words)
\item modality independent central representations (e.g., semantic represenatations of words)
\end{itemize}
In our case, our discussion will naturally focus on the latter due to assumption that semantic spaces deliver semantic representations of words.
For example, the ``Reagan" vector $\ket{r}$ from the semantic space $S_r$ illustrates a ``modality independent central representation".

Context can be conceptualised as a mental representation (overall holistic picture) of the context in which items, or events have occurred. (e.g., ``Reagan" in the context of ``Iran"). Context is also modelled as a vector of feature weights.
Following from this, context is $X$ is assumed to be represented by a ket $\ket{x}$.
In the case of the running example, the ``Iran" vector $\ket{i}$ drawn from the semantic space $S_i$ could be employed as a context vector.

Memories are associative by nature and unique representations are created by combining features of items and contexts. Several different types of associations are possible~\cite{Humphreys:Bain:Burt:89}.
The association of interest here is a two way association between a word $\ket{w}$ and a context $\ket{x}$.
In the matrix model of memory, an association between context and a word is represented by an outer product; $\ket{w}\bra{x}$.
Seeing a given word (a target) in the context of other words (cue) forms an association which probes  memory.
Observe with respect to the running example how the probe $\ket{r}\bra{i}$ embodies both the cue of the probe ``Iran" and the target ``Reagan".

In the light of the above brief digression into a matrix model of human memory, one possibility is to formalize the projector $P_x$ as the probe $\ket{w}\bra{x}$.
The object being probed is a density matrix which is not a superposition of memory traces but of semantic spaces hinged around a particular word or concept.
Equation~\ref{eqn:psem} and its density matrix equivalent (equation~\ref{eqn:density}) reflect this superposition, however in this case the traces, in their raw form, are windows of text.

In short, viewing the collapse of meaning in terms of retrieval and matching processes in memory refines the collapse equation~\ref{eqn:collapse} as follows.
Let $\ket{w}$ be a target concept and $\ket{x}$ be the context.
Firstly, collapse of meaning is characterized by projecting the probe into the memory corresponding to the state of the target word $w$. The collapse equates with retrieving a new state of meaning reflecting the change of meaning of $w$ in light of the context.
\begin{equation} \quad \quad \quad
    P_x\rho_w  = p\frac{\ket{w}\bra{x}\rho_w}{f(\xpect{x}{\rho_w}{w})} = p\rho_w^x
\label{eqn:collapse-matrix}
 \end{equation}
The probability $p$ of collapse is assumed to be a function\footnote{Further research is needed to provide the specifics of this function which will take into account issues such as decay processes in memory} of the match between the probe and the memory:
\begin{equation} \quad \quad \quad
    p = f(\xpect{x}{\rho_w}{w})
   \label{eqn:pr_collapse}
 \end{equation}
Motivating the collapse of meaning by means of the matrix model of memory introduces a deviation from orthodox QM.
After application of the  probe $\ket{w}\bra{x}$, the the state after the collapse, denoted $\rho_w^x$ is not guaranteed to be density matrix. This deviation from orthodox QM is not solely a technical issue.
It may well be that there are different qualities of probe.
For example, ``Reagan" in the context of ``Iran" would intuitively involve a projection of the global ``Reagan" semantic space onto a subspace dealing with ``Iran".
On the other hand, consider ``lion" in the context of "stone". In this case, the result after the application of the context would seem to be considerably outside the ``lion" space as a ``stone lion" does not share many of the attributes of a living one.
It would seem, then,  a projection operator is not the appropriate mechanism, but rather a more general linear operator which can project ``outside" the space.
In the latter case, equating $P_x$ with the probe $\ket{w}\bra{x}$ is arguably justified as such probes in the matrix model of memory briefly described earlier are transformations of the space, rather than projections into it. 
An alternative view is that ``stone lion" is a result of concept combination and mechanisms other than projection operators are required to suitably formalize it.
For example, Aerts and Gabora \cite{Article:05:Aerts:QMII} resort to tensor products for concept combination.
These are slippery issues requiring a clean distinction between context effects and concept combination. More research is needed to clarify these issues in relation to a logic of down below.

It remains to provide a characterization of $P_x$ as an orthodox projector as typified by  the ``Reagan in the context of ``Iran" example.
In order to do this, the senses $B = \{\ket{e_1}, \ldots, \ket{e_m}\}$ are assumed to form a basis. (The assumption here is linear independence, which is a weaker assumption than assuming the $\ket{e_i}$'s are mutually orthogonal, i.e., an orthonomal basis as is commonly seen in orthodox QM). The set $B$ represents the basis of the space $S_w$ in relation to $\rho_w$.
Let $B_x = \{\ket{x_1},\ldots, \ket{x_r}\}$ and $B_y=\{\ket{y_1}, \ldots, \ket{y_{m-r}}\}$ such that $B_x \cup B_y = B$.
The set $B_x$ is the basis of  the subspace $S_x$ due to context $X$.
The complementary space is denoted $S_y$.
By way of illustration in terms of the running example, $B_x=\{\ket{x_1},\ket{x_2}\}$ would corresponds to the two senses of ``Reagan" in the context of ``Iran" previously introduced as $\ket{e_c}$ and $\ket{e_h}$.
Though complementary spaces, $S_x$ and $S_y$ are \emph{not} assumed to be orthogonal.
Consequently, the projection operator $P_x$ is ``oblique" rather than orthogonal. Once again, this is a deviation from orthodox QM, but nevertheless faithful to the underlying intuition behind projection operators.
As stated earlier, the projection operator $P_x$ ``retrieves" the relevant senses out of the probability mixture (equation~\ref{eqn:density}), that is $P_x\ket{x_i} = \ket{x_i}$, for $x_i \in \{\ket{x_1}, \ldots, \ket{x_r}\}$.
These are the so called \emph{fixed points} of the projector $P_x$.
As a consequence, the density matrix form of the fixed points also holds as $P_x(\ket{x_i}\bra{x_i}) = (P_x\ket{x_i})\bra{x_i} = \ket{x_i}\bra{x_i}$.
This establishes that $P_x$ will retrieve the density matrix form of the relevant senses expressed in equation~\ref{eqn:density}.

$B_{n\times m}$ is an $n \times m$ matrix with columns 
\begin{eqnarray*}
[\ket{x_1}\ket{x_2}\cdots\ket{x_r}|\ket{y_1}\ket{y_2}\cdots\ket{y_{m-r}}] &= & 
[X_{n\times r}|Y_{n\times (m-r)}]
\end{eqnarray*}
The projection operator $P_x$ retrieves those fixed points relevant to the context. All other senses are projected onto the zero vector $\ket{0}$:
\begin{eqnarray*}
P_xB &= & P_x[X|Y] \\
         & = & [P_xX|P_xY]\\
         & = & [X|0]
\end{eqnarray*}
For the case $m=n$, the matrix $B$ has an inverse $B^{-1}$ so the makeup of the required projection operator is given by:
\begin{equation} \quad\quad\quad
P_x  =  [X | 0] B^{-1} 
         = B \left ( \begin{array}{cc}
                         I_r & 0 \\
                         0 & 0 
                         \end{array}
                \right ) B^{-1}         
\end{equation}  
With an eye on operational deployment on a large scale,        
a simple algorithmic construction of $P_x$ is based on the intuition that those senses which are not orthogonal to the cue  should be retrieved from the linear combination of $m$ senses (equation~\ref{eqn:density}):
\begin{eqnarray*}
 B_x = \{\ket{e_i} | \xpect{x}{\rho_i}{x} > 0, 1 \leq i \leq m\}
\end{eqnarray*}
(Recall that $\rho_i = \ket{e_i}\bra{e_i}$).
In terms of the running example, $B_x = \{\ket{e_c}, \ket{e_h}$, the two senses relevant to ``Reagan" seen in the context of ``Iran".

The scalar $\xpect{x}{\rho_i}{x}$ decomposes as follows:
\begin{eqnarray*}
\xpect{x}{\rho_i}{x} & = & \bra{x}(\ket{e_i}\bra{e_i})\ket{x} \\
                              & =  & (\scprod{x}{e_i})^2 \\
                              & =  & \cos^2 \theta_i \\
                              & =  & a_i
\end{eqnarray*}
where $\cos \theta_i$ is the angle between $\ket{x}$ and $\ket{e_i}$.
In the second last line the equivalence between Euclidean scalar product and cosine was employed due to the vectors being normalized to unit length.
This value reflects how much the given sense is being activated to the level $a_i$ by the cue $\ket{x}$. 
Stated otherwise, $a_i$ reflects the strength with which the sense $\rho_i$ is aligned with the cue $\ket{x}$.
All senses $\ket{e_i}$ in the basis $B_x$ will have a positive activation value $a_i$.
By appropriately scaling the values $a_i$,
the effect of projector $P_x$ can now be expressed as a probability mixture:
 \begin{equation} \quad \quad \quad
    P_x\rho_w  = p_1\rho_1 + \ldots + p_r\rho_r
   \label{eqn:Pxweight}
  \end{equation}
  where $\rho_i = \ket{e_i}\bra{e_i}$, for all $\ket{e_i} \in B_x$ and $p_1 + \ldots + p_r=1$.
The import of the last equation is that the effect of the projector $P_x$ results in a density matrix.

\subsection{The probability of collapse}

It is illustrative to examine how in the light of the running example the scalar value resulting from the matching process determines the probability of collapse (equation~\ref{eqn:pr_collapse}). First, the effect of the cue ``Iran" via the context vector $\ket{i}$ is shown.
The ``memory" to be probed derives from the target ``Reagan" and is denoted by the density matrix $\rho_r$.
\begin{eqnarray*}
\lprod{i}{\rho_r} & = & \lprod{i}{(p_1\rho_1 + \ldots + p_m\rho_m)} \\
                         & = & p_1\lprod{i}{\rho_1} + \ldots + p_m\lprod{i}{\rho_m}
\end{eqnarray*}
Recall that each of the $m$ constituent density matrices $\rho_i$ derives from a particular sense of ``Reagan" denoted $e_i$. Therefore the previous equation can be written as,
\begin{eqnarray*}
\lprod{i}{\rho_r} & = & p_1\lprod{i}{(\ket{e_1}\bra{e_1}}) + \ldots + p_m\lprod{i}{(\ket{e_m}\bra{e_m}}) \\
                         & = & p_1(\scprod{i}{e_1})\bra{e_1} + \ldots + p_m(\scprod{i}{e_m})\bra{e_m} \\
                         & = & p_1 \cos \theta_1 \bra{e_1} + \ldots + p_m \cos \theta_m \bra{e_m}
\end{eqnarray*}
The salient facet of the last line is  those senses that are not orthogonal to the context vectors will be retrieved ($\cos \theta_i > 0$)and will contribute to the probability of collapse.
This accords with the intuitions expressed in the previous section.
In the running example, these senses were denoted $\ket{e_c}$ and $\ket{e_h}$. So,
\begin{eqnarray*}
\lprod{i}{\rho_r} & = & p_c \cos \theta_c \bra{e_c} + p_h \cos \theta_h \bra{e_h}
\end{eqnarray*}

A second aspect of the matching is post multiplying with the target vector ``Reagan", denoted $\ket{r}$:
\begin{eqnarray*}
\rprod{(p_c \cos \theta_c \bra{e_c} + p_h \cos \theta_h \bra{e_h})}{r} & = & p_c \cos \theta_c(\scprod{e_c}{r}) +  p_h \cos \theta_h (\scprod{e_h}{r}) \\
   & = & p_c \cos \theta_c \cos \psi_c + p_h \cos \theta_h \cos \psi_h \\
   & = & p_c m_c + p_h m_h
\end{eqnarray*}
The angles $\cos \psi$ reflects how strongly the the sense correlates with the given target. It can be envisaged as a measure of significance of the given sense with the target $\ket{r}$.
The scores due to matching of the probe with memory are reflected by the scalars $m_c$ and $m_h$. These are modified by associated probabilities of the respective senses.
Finally, the two terms are added to return a single scalar.
The probability of collapse is assumed to be a function of this value.

\subsection{Summary}

The preceding development has centred around providing an account of the collapse of meaning in the light of context.
It is important that the formalization rests on non-orthogonal density matrices, which is in contrast to the orthogonal approach used in the SCOP model \cite{Article:05:Aerts:QMII}.
The approach presented here draws inspiration from a cue which probe human memory and describes collapse of meaning in terms of memory cues. 
The notion of a ``probe" is not foreign to QM. The most useful
probes of the various wave functions of atoms and molecules are
the various forms of spectroscopy. In spectroscopy , an atom or
molecule  starting with some wave function (represented by a
density matrix)  is probed with light, or some other particle. The
light interacts with the molecule and leaves it in another state.
This process is analogous to the probing of memory just described.
Chemical physics also shares another similarity with our account
in the sense that the underlying density matrices cannot be
assumed to be orthogonal. Nonorthogonal density matrix
perturbation theory has arisen to deal with nonorthogonal density
matrices and may turn out to be a relevant area for formalizing
additional aspects of a logic of ``down below". The analogy should
be mindfully employed, however. Human memory is a vast topic
abundant with texture and nuance, not to mention strident debate.
However we feel investigations into the memory literature can bear
further fruit in relation to a QM inspired account of a logic of ``down below".
The matrix model of memory described above has been
extended to provide an account of analogical
mapping~\cite{Wiles:Halford:1994}. In our opinion, it is
reasonable to assume that analogical reasoning has  roots in
subsymbolic logic. Dunbar~\cite{Article:99:Dunbar:Abduction}
concludes from cognitive studies that scientists frequently resort
to analogies when there is not a straightforward answer to their
current problem. Therefore, analogical reasoning sometimes plays a
crucial role in hypothesis formation which is fundamental to
abduction\cite[Chapter 7]{Gabbay:Woods:2005d}.
Reasoning, then, becomes highly confounded with memory processes.
Consider the ``Tweety" example described earlier. When one learns
that ``Tweety is a penguin", it is debatable whether any reasoning
takes place at all. We would argue that the example can be
explained in terms of probes to memory and the associated dynamics
of defaults emerge out of context effects. We have argued such
probes bear a striking similarity to quantum collapse.

\remove{
The probability of the collapse is given by $\Pr(\ket{e_i}) = tr(\rho_wP_x)$.
By way of illustration, let $P_x = \ket{w}\bra{x}$ and let $\rho_w  = p_1\ket{w_1}\bra{w_1} + \ldots + p_m\ket{w_m}\bra{w_m}$.
Then,
\begin{eqnarray*}
tr(\rho_wP_x) &= p_1\scprod{w_1}{x}^2 + \ldots + p_m\scprod{w_m}{x}^2
\end{eqnarray*}
This is the weighted sum of the squares of $\cos \theta_i$ where $\theta_i$ is the angle that $\ket{x}$ makes with the $i-th$ sense of the concept $\ket{w_i}$~\cite[p. 14]{Book:04:Keith:QM}.
This clearly shows how the resultant probability is being computed from the geometry of the space.

\comment{PB}{ The following equation from orthodox QM gives the state after collapse. This is almost what we are saying for real valued Hilbert spaces - any differences need to be discussed and explained}

In real valued state spaces\footnote{For the corresponding equation in complex Hilbert spaces, see~\cite{Article:05:Aerts:QMII}}
, the state $\rho_w^x$ after collapse is given:
\begin{equation} \quad \quad \quad
   \rho_w^x  = \frac{\rho_wP_x}{tr(\rho_wP_x)}
   \label{eqn:poststate}
 \end{equation}

\comment{PB}{Problem  : stone lion example? - result takes us out of the lion space, hence axiom 3 of QM doesn't hold}

\comment{PB}{How do  we go from here into QL and what will this QL look like?}
}

\section{Quantum Logic and Conceptual Generalization}
\label{QL-generalization-sec}

A proposal for reasoning at the subsymbolic level must give an account
for how conceptual structures may arise from perceptual
observations. For example, in {\it Word and Object}, Quine \cite[p. 25]{Quine:1960}
famously challenged philosophers to give an account for how a hearer
might reliably deduce that a speaker who utters the word ``gavagai"
upon seeing a rabbit actually means ``rabbit", instead of ``part of a
rabbit", or a member of some other class such as ``rabbit or guppy", or
even ``rabbit or Reagan". In other words, how might a conceptual logic
give rise to a recognition and representation of natural kinds, in
such a way that this logic is cognitively beneficial?

It is known that some logics are more amenable to inductive learning
than others, and that direct adherence to the Boolean distributive law
effectively prevents the sort of smoothing or closure operations that
may lead to the formation of natural kinds (see\cite{Article:04:Widdows:QM}).
For example, since Boolean logic is modelled
on set theory and the union of the set of rabbits and the set of frogs
is a perfectly well-formed set, the concept ``rabbit or guppy" is as
natural as the concept of ``rabbit" in Boolean logic. At the other
extreme, a single-inheritance taxonomic logic (based, for example, on
phylogenetic inheritance) may overgeneralize by assuming that the
disjunction of  ``rabbit" and ``guppy" must be the lowest common
phylogenetic ancestor ``vertebrate". This would lead also to
unfortunate consequences, such as the presumption that, since a rabbit
makes a good pet for a child and a guppy makes a good pet for a child,
any vertebrate makes a good pet for a child.

Compared with the discrete extreme of Boolean classification, and the
opposite extreme of a single-inheritance taxonomy, the vector lattice
of quantum logic presents and attractive middle ground. There are
distinctly well-formed concepts represented by lines and planes, there
is a natural closure or smoothing operator defined by the linear span
of a set, and there is a scope for multiple inheritance (since a line
is contained is many different planes and an $m$-dimensional subspace
is contained in many $m+n$-dimensional subspaces). Some practical
evidence for the usefulness of the linear span as a disjunction of two
concept vectors was provided by the experiments in
\cite{widdows2003negation}, in which the removal of a pair of concepts
using negated quantum disjunction proved greatly more effective than
Boolean negation at the task of removing unwanted keywords and their
synonyms.  The
argument that projection onto subspaces of a vector space can be used
as a solution to the age-old problem of learning from incomplete
experience has been made one of the mainstays of Latent Semantic
Analysis, by \cite{Article:97:Landauer:LSA} and others.

\begin{figure}
\begin{center}
\epsfig{file=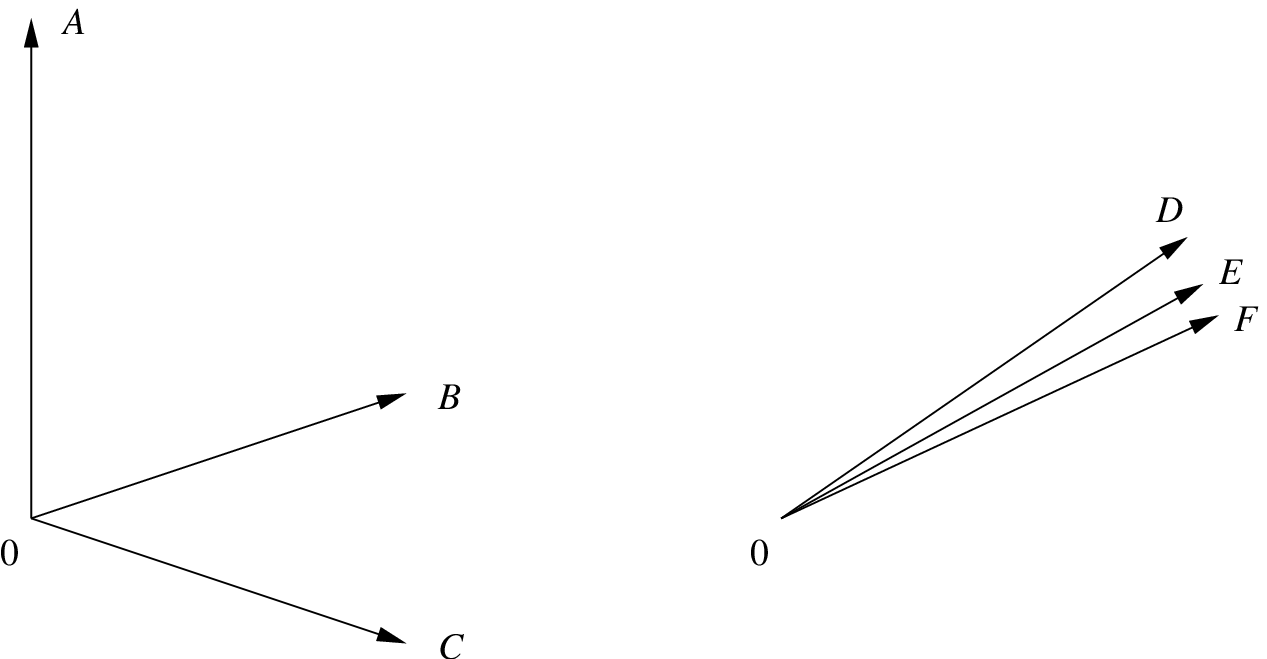,angle=270,width=4in}
\end{center}
\caption{Orthogonal vectors in 3-space compared with 3 similar vectors.}
\label{3d-spans}
\end{figure}

It should also be noted that the use of a pure quantum logic for
concept generalization in semantic space leads to problems of its own,
as one would expect with any attempt to apply such a simple
mathematical model to a wholesale description of language. In
particular, quantum disjunctions may often overgenerate, because of
the nature of the linear independence and the operation of taking the
linear span. In practice, vectors that are very close to one another
in semantic space may still be linearly independent, and will thus
generate a large subspace that does not reflect that fact that the
vectors were in fact drawn from a small region of this subspace. This
danger is illustrated in Figure \ref{3d-spans}, which depicts two
groups of three vectors in a 3-dimensional vector space. In the left
hand picture, the vectors $A$, $B$ and $C$ are orthogonal and can be
used to generate the whole of the space. The vectors $D$, $E$ and $F$,
far from being orthogonal, have high mutual similarity. However, since
these vectors are still linearly independent, they can still be used
to generate the whole of the space. In other words, the quantum
disjunctions $A\vee B \vee C$ and $D\vee E \vee F$ are identical. This
seems quite contrary to intuition, which would suggest that the
concept $D\vee E \vee F$ should be much more specific than the concept
$A\vee B \vee C$.
A practical drawback of this overgeneration is that
a search engine that used quantum disjunction too liberally would be
likely to generate results that would only be judged relevant by users
willing for their queries to be extrapolated considerably.

There is a natural way to fix this problem in the formalism, and it
bears an interesting relation to the observation that non-orthogonal
vectors and subspaces give rise to subtly related non-commuting
density matrices.
In the diagram, the vector $E$ lies nearly but not
quite upon the line from $D$ to $F$. To simplify the description of
the local situation, a reasonable approximation would be to represent
$E$ by its projection onto the subspace $D\vee F$. This would amount
to making the assertion ``$E$ is between $D$ and $F$'', which might
not be {\it exact}, but from a human standpoint is certainly {\it
  true}. To generalize from this example, it would be reasonable to
say that a vector $B$ can be {\it approximately derived from} a set
${A_1, \ldots, A_j}$ if distance between $B$ and the projection of $B$
onto $A_1\vee \ldots \vee A_j$ is small. Defining `small' in practice
is a subtle challenge, and to some extent is in the eye of the
beholder --- to some, {\it Liszt} is adequately represented as being
somewhere between {\it Beethoven} and {\it Brahms}, but to some, his
music has special qualities independently of both.

Such a discussion suggest that one of the requirements for a working
quantum logic of semantic space is the ability to model automatically
the ``natural dimension'' of a sample of points. This problem will
have many variants, and different solutions will be appropriate for
different users. However, there are some general techniques such as
the Isomap algorithm \cite{tenenbaum2000isomap} that provide
dimensional decompositions of this sort, even for samples of points
taken from nonlinear submanifolds of vector spaces. From a cognitive
point of view, such a dimensional simplification is to be expected and
indeed preferred. From microscopic observation and subsequent progress
in chemistry and physics, we know that the surface of a wooden
tabletop is a complex 3-dimensional structure, which may have a
detailed fractal surface and according to some physical theories may
consist of particles that need several more dimensions to be
represented correctly. However, even to those humans who are well
versed in such scientific precision, the tabletop is for all practical
purposes a 2-dimensional structure, and you can certainly describe the
whereabouts of any perceptual object on the table at a relevant scale
of reference by giving two coordianates.

The challenges for adapting the vector space model to describing
semantics and perception do not end here, of course. Many of the
vector space axioms (such as the underlying assumption that vectors
form a commutative group under addition) are seriously off the mark when
viewed from a cognitive perspective.
The purpose of this discussion
is not to convince the reader that these problems have been completely
solved, but that the immediate drawbacks of a naive implementation of
quantum logic and concept formation in semantic space can be
anticipated by a more careful consideration of the cognitive and
logical goals of the system, whereupon plausible solutions can be
found using existing mathematics.

\section{Summary and Conclusions}
A logic that is shaped by the empirical make-up of reasoning
agents is subject to the same experimental challenges and
limitations that affect the investigation of human subjects quite
generally. The interior of the atom is, in well-known ways,
difficult to access, but the interior of the reasoning agent
throws up accessibility difficulties of an entirely different
order. Experimental psychology, to take the most obvious example,
has had to learn how to flourish despite the collapse of
behaviourism and introspectivism. A great part of its success,
such as it is, is owed to the skill with which it organizes its
theoretical outputs around strongly plausible conjectures. In a
rough and ready way, conjecturing is what one does in the absence
of observation. In this regard, we are reminded of the grand
conjecture with which Planck launched quantum theory itself, an
idea whose immediate import in 1900 was the unification of the
laws of black body radiation. In taking a quantum approach to the
logic of reasoning down below, two sources of conjecture merge. In
probing the down below, we conjecture in ways that incorporate the
conjectures of quantum mechanics. Of course, by now there is ample
empirical confirmation of quantum theory, as well as encouraging
empirical support for cognitive psychology in some of its
manifestations, but neither of those desirable outcomes would have
been possible without the founding conjectures around which the
original theories organized themselves. It is the same way with
the logic of down below. A reliable empirical understanding of it
has no chance of occurring spontaneously. It must be preceded by
theoretical speculation. Eddington once quipped that theories are
``put-up jobs'', anticipating Quine's crack that theories are
``free for the thinking up''. These, of course, are jokes. The
fact is that the practice of scientific conjecture is respectable
to the degree that it conforms to the canons of abductive logic.
One of the marks of an abductively successful conjecture is its
{\em narrative coherence} with what is known observationally about
the subject in question.
\cite{Thagard:1989},\cite{Gabbay:Woods:2005d}. Smooth narratives
identify possible scenarios. This is how we find ourselves
positioned here. We have sketched what we take to be a coherent
narrative of the quantum character of reasoning down below. To the
extent that we have succeeded in this, we have outlined a possible
theory for such reasoning. What remains now is to sort out ways in
which the theory might be made responsive to observational test.
Initial steps in this direction have been made in the
realm of text-mining and search, a field which benefits from the
comparative ease of empirical measurement and hypothesis
testing. Whether the theory provides a truly useful model of
cognitive processes will require different observational methods. 

\remove{
\section{Things to put somewhere}

\begin{itemize}
\item make the connection between G\"{a}rdenfors' regions and subspace
\end{itemize}
}
\subsubsection*{Acknowledgments}
The first author acknowledges National ICT Australia which is funded by the Australian Government's Department of Communications, Information Technology, and the Arts and the Australian Research Council through \emph{Backing Australia's Ability} and the ICT Research Centre of Excellence programs.
The third author thanks Dov Gabbay and Kent Peacock for helpful
advice, and, for its financial support, the Engineering and
Physical Sciences Research Council of the United Kingdom. Thanks to Ian Turner for helpful comments of a technical nature.


\end{document}